\documentclass[12pt]{article}
 
\usepackage{epsfig}
\begin{document}

\baselineskip=14pt plus 0.2pt minus 0.2pt
\lineskip=14pt plus 0.2pt minus 0.2pt

\newcommand{\be}{\begin{equation}}
\newcommand{\ee}{\end{equation}}
\newcommand{\bea}{\begin{eqnarray}}
\newcommand{\eea}{\end{eqnarray}}
\newcommand{\da}{\dagger}
\newcommand{\dg}[1]{\mbox{${#1}^{\dagger}$}}
\newcommand{\hlf}{\mbox{$1\over2$}}
\newcommand{\lfrac}[2]{\mbox{${#1}\over{#2}$}}
\newcommand{\scsz}[1]{\mbox{\scriptsize ${#1}$}}
\newcommand{\tsz}[1]{\mbox{\tiny ${#1}$}}

\thispagestyle{empty}

\begin{flushright} 
~
\\ ~ LA-UR-02-294 
\\
\end{flushright} 
\begin{center}

\large{\bf A MISSION TO TEST THE PIONEER ANOMALY\footnote{
This essay received an honorable mention in the Annual Essay
Competition of the Gravity Research Foundation for the year 2002
  --- Ed.}}

\vspace{0.5in}

\normalsize
\bigskip 

JOHN D. ANDERSON,$^a$ MICHAEL MARTIN NIETO,${^b}$ and SLAVA G. 
TURYSHEV$^c$ \\

\normalsize
\vskip 15pt

$^{a,c}${\it Jet Propulsion Laboratory, California Institute of  Technology,\\
Pasadena, CA 91109, U.S.A.} \\ 
\vskip 5pt
${^b}${\it Theoretical Division (MS-B285), Los Alamos National Laboratory,\\
University of California,  Los Alamos, New Mexico 87545, 
U.S.A.}\footnote{Email
addresses: {\tt john.d.anderson@jpl.nasa.gov},   {\tt mmn@lanl.gov}, {\tt
turyshev@jpl.nasa.gov}} 
\\


\vspace{0.5in}
\bigskip 

\baselineskip=.33in

\end{center}

\baselineskip=.33in

Analysis of the radio tracking data from the Pioneer 10/11 spacecraft
has  consistently indicated the presence of an anomalous small Doppler
frequency drift.   The drift can be interpreted  as
being due to a constant acceleration of
$a_P= (8.74 \pm 1.33) \times 10^{-8}$  cm/s$^2$  directed {\it towards}
the Sun.   Although it is suspected that there is a systematic origin
to the effect, none has been found.  The nature of this anomaly has
become of growing interest in the fields of relativistic cosmology, astro- and
gravitational physics as well as in the areas of spacecraft design and
high-precision navigation.
We present a concept for a designated deep-space mission to test the 
discovered anomaly.  A number of critical requirements and design
considerations for such a mission are outlined and addressed.   

\vspace{0.3in}


\newpage

\baselineskip=.33in

\section{The Pioneer Missions and the Anomaly}

The Pioneer 10/11 missions, launched on  2 March  1972 (Pioneer 10) and 4
Dec 1973 (Pioneer 11), were the  first to explore the outer solar system
\cite{science}.   After Jupiter and (for Pioneer 11) Saturn
encounters, the two spacecraft followed escape hyperbolic orbits near  the
plane of the ecliptic to  opposite sides of the solar system. Pioneer 10
eventually became the first man-made object to leave the solar
system.  

Pioneer 10's radio signal is weakening.  Despite this, the Deep
Space Network (DSN) is still able to deliver navigational data from
distances $\sim$80 AU.  Indeed, on the  30th anniversary of its launch, 
2 March 2002, the Madrid DSN station received a return radio signal
from Pioneer 10 at 22:47 CET.  This was 22 h 06 m after the 
uplink signal was sent from the Goldstone, CA, DSN station.   

By 1980, when Pioneer 10 was at   a distance of $\sim 20$ AU from the
Sun, the acceleration contribution from solar-radiation-pressure on
Pioneer 10 (directed {\it  away} from the Sun) decreased to  
$< 5 \times 10^{-8}$ cm/s$^2$.  At that point the navigational data 
began to clearly indicate  the presence of an anomaly in the Doppler
navigational data, which was later interpreted as  a constant
acceleration, $a_P$,  directed {\it toward} the Sun \cite{anderson}.

Recently, we published a detailed study of the Pioneer
anomaly, which  used the existing Pioneer 10/11 Doppler
data from 1987.0 to 1998.5. We specifically addressed all possible sources
for a systematic origin for the detected anomaly.  Our conclusion was 
that, even after all {\it known}   systematics are accounted for, there
remains an anomalous  acceleration signal of
$a_P= (8.74 \pm 1.33) \times 10^{-5}$  cm/s$^2$  directed {\it towards}
the Sun \cite{pioprd}. 

We emphasize {\it ``known''} because we
must admit that the most  likely cause of the effect is some as yet not
understood systematic generated by the spacecraft themselves,  perhaps
caused by excessive heat or propulsion gas leaks. 
But neither we nor others with spacecraft or navigational
expertise have been able to find it \cite{pioprd}.   

Further, due to its different mission and spacecraft designs, as well
as its proximity to the Sun, the use of the Cassini spacecraft to test
for the anomaly proved to be impractical.  A number of alternative
ground-based verifications of the anomaly were also considered; for
example, using Very Long Baseline Interferometry (VLBI)  astrometric
observations.  However, the trajectory of Pioneer 10,
with a small proper motion on the sky, makes it presently impossible to
accurately isolate the anomalous Sun-ward acceleration.

Therefore, we strongly argue that the time has come to  consider a
new deep-space experimental test of this intriguing effect.  


\section{ The Mission}

When considering any space mission one needs to address a number of 
important issues, such as 
(i) the scientific justification for the mission objectives;
(ii) the mission configuration and design requirements; and 
(iii) the overall construction, launch, and ground operations cost.  
The scientific justification for our mission is clearly outlined 
above.  

The cost would be the most constraining factor.  Almost any deep-space 
mission would now cost on the order of M\$300-500 \cite{cost}. 
Therefore, a test of the Pioneer effect 
might best be considered as a relatively small part of another dedicated
mission whose objective is to study the boundaries of the solar system.   

First, one needs to be at a distance 
greater than 20 AU to be able to best distinguish any effect 
from solar radiation pressure and other near-solar systematics.  
Therefore, a very energetic rocket would be helpful. We
observe that  the Russian Proton rocket is an intriguing possibility.   
Indeed, this might be a useful option for international collaboration 
and to hold down the cost to NASA \cite{bestworld}.  Further, NASA
has renewed interest in nuclear rockets \cite{nuc}. This might enable 
faster missions to the outer solar system.

\begin{figure}[h]
 \begin{center}
\noindent    
\psfig{figure=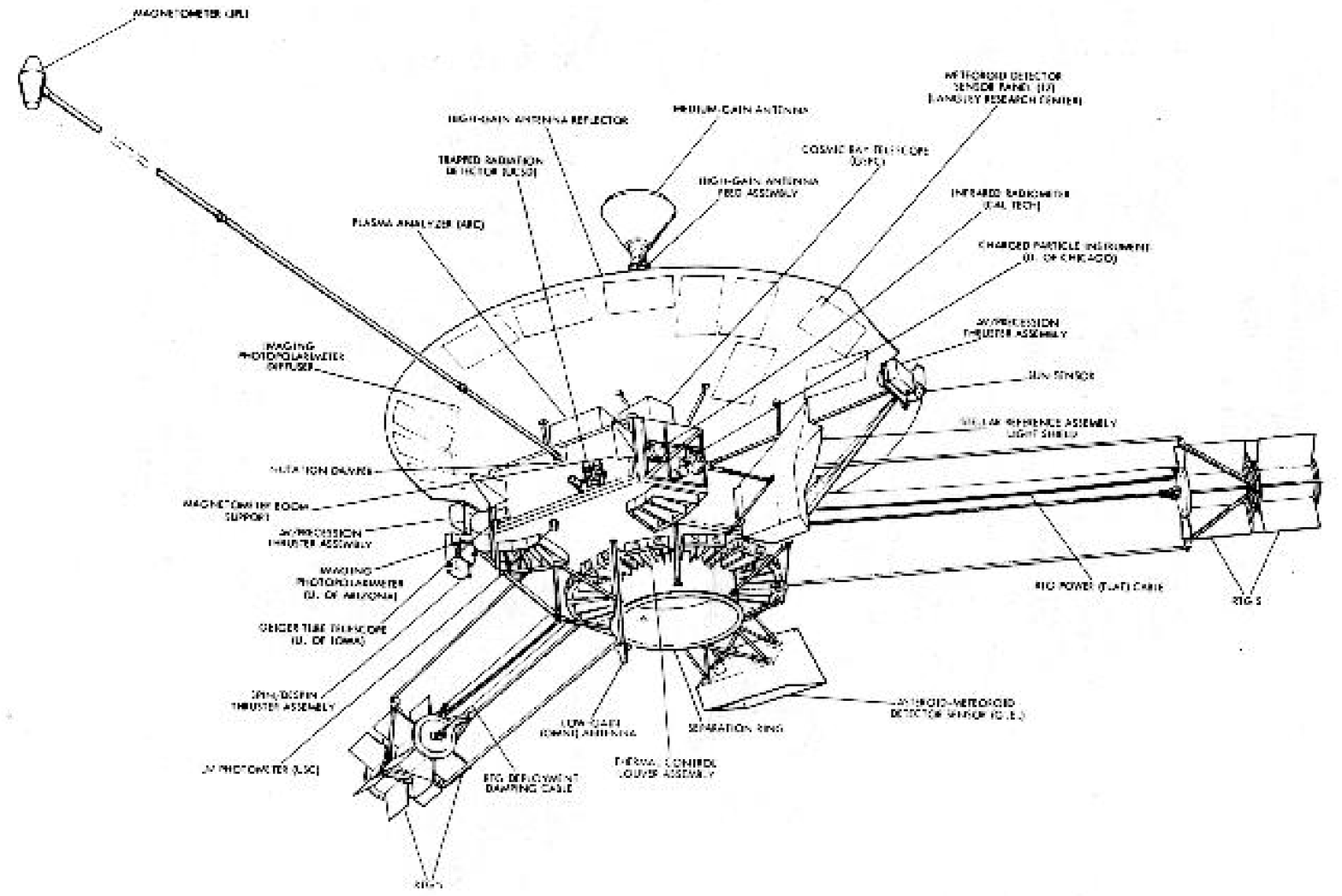,width=155mm}
\end{center}
  \caption{A drawing of the Pioneer spacecraft.  
 \label{fig:trusters}}
\end{figure}

Non-gravitational forces acting on
spacecraft are common and they can cause problems for precision space
navigation. So, once in deep space one needs to have a
spacecraft with very small or else well-understood systematics.  
Therefore, an anomaly test could well impose stringent design constraints.    
To better understand what is needed, it is useful to  keep in mind 
the design of the Pioneer craft and to understand what made them work so well.
(See Figure 2 of
Ref. \cite{pioprd}, the present Figure \ref{fig:trusters}.) 


Among the most important features of the Pioneers were 
\cite{pioprd}: (i) simple,
spin-stabilized attitude
controls; (ii) on-board nuclear power sources, (iii)
a well-understood thermal control system; 
(iv) extensive navigational coverage with high accuracy Doppler 
tracking; and (v) hyperbolic escape-orbit trajectories. 
So, in light of our experience studying the Pioneer anomaly,
consider the main requirements on a spacecraft design. 

\paragraph{Attitude control:} 
For navigational purposes Pioneer spacecraft are much
simpler than any other spacecraft, including Voyager, Galileo, and
Cassini.  The two Pioneers are simple spinners and thus they
have no continuous jetting of attitude control gas. Moreover, 
in deep space they require
only a single maneuver every few months or so to correct for the effect
of proper motion. That is one of the main reasons they are so well
tracked.  

To be a simple spinner, the craft with its equipment will have to be
moment-of-inertia balanced about the main antenna axis. This can be aided
by having the Radioisotope Thermoelectric Generators (RTGs), which
generate the electrical power, on extended booms that are deployed after
launch.

However,  other mission objectives might necessitate a  3-axis stabilized
spacecraft.  If so, then in order to achieve comparable navigational
accuracy  one would need to develop and fly long-lasting accelerometers, 
very precise fuel gages, and well-calibrated thrusters.   

Finally, dual stabilization might be used;  3-axis near encounters and
spin stabilized on cruise, as was done for Galileo.  
In any event, one wants spin-rate control and/or accelerometers 
that would yield measurements accurate to the level of the 
Pioneer navigational error, $\mathcal{O}(10^{-9})$ cm/s$^2$ 
[in other units, 
$\mathcal{O}(10^{-12})$ g~$= \mathcal{O}(10^{-3})~\mu$Gal].

\paragraph{On-board power system:}
RTGs are the only viable choice for deep space power,  so   
international cooperation might again be useful. 
Due to environmental politics (recall the Cassini Earth-flyby 
furor) the USA no longer makes $^{238}$Pu for RTGs.  
This could change with NASA's new nuclear initiative \cite{nuc}, but
for now, Russia is the only source.  

\paragraph{Heat rejection and thermal control:}
The RTGs bring up the other main systematic in deep space, thermal 
emission generated by the spacecraft's power system.  One reason
the Pioneer RTGs were placed on booms was fear of gamma radiation damage
to  the spacecraft electronics and surface.  This turned out not to be a
problem but the placement was serendipitously lucky. The RTGs, with 
$\sim2,\!500$
W of heat, were placed where they would have little thermal effect on the
craft.   (The Pioneer effect could be caused by only 63 W of directed
power radiating from the 241 kg craft.)  The rotation of the craft and
the  RTG fin structures were designed to radiate symmetrically fore-aft,
with much less heat radiated in the direction towards the craft. The same
concept should be used for this mission, with perhaps shielding of the 
craft to prevent anisotropic heat  reflection. 

The electrical power in the equipment and instrument compartments
must be radiated so as to not cause an undetected systematic.  For the
Pioneers the central compartment was surrounded by insulation with
louvers aft to let out excess heat early in the mission, and to retain
heat later on when the electrical power was less. The electrical power
degrades faster than the radioactive decay because the thermoelectric
devices deteriorate. 

For this mission, the louvers should be on the side of the compartment so 
they will radiate in an axially symmetric manner as the spacecraft
rotates.  The top and bottom of the compartment should also be insulated
to further minimize the heat transfer and reflection. The booms
connecting the compartment to the antenna as well as the booms to the
RTGs should be thermally isolated, either with insulators in the
structure or by using appropriate materials in the construction.  As good
as possible a priori thermal models should be created and test-stand
measurements and calibrations should be made.

\paragraph{Communications:}
Even with all systematics known, no good data is possible without good
navigation. This implies the use of both Doppler and range data. 
The Doppler tracking, which measures the velocity of the craft, should be
done at two frequencies, say  X-band and Ka-band. The two frequencies are
useful to correct for dispersive media effects and will allow precise 
calibration of plasma systematics.  But the Doppler technique only 
indirectly measures distance to the craft, by integrating the measured
Doppler velocity from known initial conditions.  Range itself is a
time-of-flight measurement.  This is done by phase modulating the signal
and timing the return signal, which was transponded at the craft.  
As such, it gives the distance to the spacecraft directly.

\paragraph{Three-dimensional tracking:}
Having both Doppler and range would allow a very precise orbit to be
determined, especially if VLBI were used.  Indeed. one might be able
to obtain good three-dimensional acceleration data.  
This latter would be very desirable for detailed acceleration 
anomaly searches.  One would expect that an internal systematic
would be directed along the craft spin axis, an anomalous new force 
would be directed towards the Sun, an external drag force would be 
almost along the velocity vector, and a time acceleration would be 
directed towards the Earth.  Having three-dimensional tracking might
allow a differentiation to be made from among these four directions. 

\paragraph{Turning the spacecraft around:}
We also propose an experiment which would clearly determine how
much, if any, of an anomaly were due to systematics.  Suppose one had
an additional antenna in the forward direction, appropriate care taken
of the mounting of the craft to the launch vehicle.  (One would also
continuously transmit from both antennas to reduce the radio-power
systematic.)  Then, aided by Sun and star sensonrs, if one rotated
the spacecraft by 180$^\circ$ so that the forward
antenna faced the Earth, any systematic
would be in the opposite direction whereas an acceleration due to an
exterior force would not.  A very similar rotation was actually 
performed on Pioneer 10 soon after launch, the Earth acquisition 
precession.  For a craft like the Pioneers
such a maneuver could be done in about two hours and take about 0.5 kg of
fuel \cite{lozier}.  

Finally, if the Pioneer anomaly is due either to ``normal gravity,'' 
or to ``non-gravitational force'' but not to systematics, it
can be measured by Doppler and range.  
But accelerometers, used as feed back to control non-gravitational  
forces, would distinguish between the two types of forces.  
(Modified inertia \cite{mond},  
violating the strong equivalence principle, might also be addressed.) 


\section{Summary}

Since future space missions will require accurate navigation and/or
positioning, 
it is important that we gain an understanding of the Pioneer anomaly. 
For instance, the Space Interferometry Mission (SIM) and the Laser
Interferometric Space Antenna (LISA) [and possibly even a mission to
Pluto and the Kuiper Belt] want navigation to a precision less that
that which would be caused by the Pioneer anomaly.   In
particular, if the  Pluto/Kuiper mission goes, it would be fortunate if
the  craft were designed so as to be able to repeat the Pioneer
measurement beyond 20 AU. 

It is our personal hope that the next generation of deep space missions
will utilize a much greater navigational accuracy and precision. Then, 
independent of whether the Pioneers measured a systematic effect
generated by the spacecraft or (unlikely as it is)  a manifestation of
``new physics,'' an experiment to test the result would be important and 
should be done \cite{esa}. If, as is probably the case, 
the anomaly is due to some systematic, understanding
this will greatly aid  future mission design and 
navigational programs.    But if, on the other hand, there is something
unknown going on, the implications are obvious.  


\section*{Acknowledgments}

We first express our deep appreciation of the insights 
obtained from  our colleagues in the study of the Pioneer anomaly, 
Philip A. Laing, Eunice L. Lau, and the late Tony S. Liu.  
for this work, David Lozier \cite{lozier} of NASA/Ames was
particularly helpful. 
The work of J. D.A. and  S.G.T was performed at the
Jet Propulsion Laboratory, California Institute of Technology, under
contract with the  National Aeronautics and Space Administration.  
M.M.N. acknowledges support  by the U.S. DOE.





\begin{thebibliography}{99}
\bibitem{bestworld}  In the best case, probes would be sent in
three directions: parallel to the Solar
system galactic motion, towards the galactic center,
and perpendicular to the ecliptic plane.
\end{thebibliography}
\end{document}